\documentclass[11pt,twoside]{article}

\usepackage{asp2006}
\usepackage{epsf}
\usepackage{lscape}
\usepackage{natbib}

\def \degr{^\circ}

\begin{document}

\title{Modeling the Polarization of Dusty Scattering Cones in
  Active Galactic Nuclei}

\author{Ren\'e W. Goosmann}
\affil{Astronomical Institute, Academy of Sciences, Prague, Czech Republic}

\author{C. Martin Gaskell}
\affil{Dept. Physics \& Astronomy, Univ. of Nebraska, Lincoln,
NE, USA}

\author{Masatoshi Shoji} \affil{Astronomy Department, Univ. of Texas,
Austin, TX, USA}

\begin{abstract}
We have used the {\sc Stokes} radiative transfer code, to model
polarization induced by dust scattering in the polar regions of Active
Galactic Nuclei (AGN). We discuss the wavelength-dependence of the
spectral intensity and polarization over the optical/UV range at
different viewing angles for two different types of dust: a Galactic
dust model, and a dust model inferred from extinction properties of
AGN. The {\sc Stokes} code and documentation are freely available at
http://www.stokes-program.info/.
\end{abstract}

Spectropolarimetry delivers important results helping to explain the
geometry of AGN \citep[e.g.][]{antonucci2002}. The spectropolarimetric
properties of individual objects are generally complicated and need to
be disentangled by accurate modeling. The new Monte-Carlo radiative
transfer code, {\sc Stokes}, is designed to simulate polarization by
scattering in a wide variety of astrophysical contexts. The program
allows the user to define different emission and scattering geometries
and solve the radiative transfer in 3D. It takes into account
polarization induced by scattering off free electrons and dust
grains. The dust composition includes carbonaceous and siliceous
grains and the grain size distributions parameterized by truncated
power laws. The code measures light travel times and thus can be used
to model polarization reverberation mapping \citep{shoji2005}.

\begin{figure}[t]
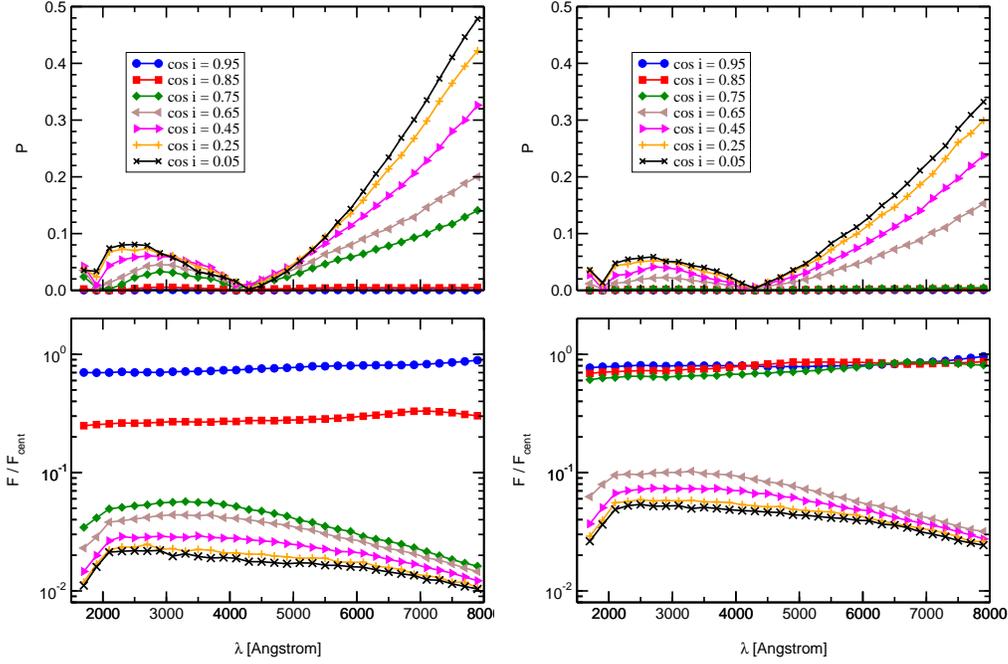

  \vskip 0.2cm \centering \epsfxsize=6.6cm
  \epsfbox{goosmann-poster-pol-fig1.eps} \epsfxsize=6.6cm
  \epsfbox{goosmann-poster-pol-fig2.eps}
  \caption{Modeling a dusty double cone of half-opening angle
  $\theta_C = 30\degr$ (left) and $\theta_C = 45\degr$ (right). Top:
  polarization, $P$. Bottom: the fraction, $F/F_*$, of the central
  flux, $F_*$, seen at different viewing inclinations, $i$.}
  \label{fig:res}
\end{figure}

Here we use {\sc Stokes} to model the polarization expected from
scattering in dusty, centrally-illuminated double cones of AGN. We
follow the same procedure as in \citet{goosmann2007}. The optical
depth of one cone along the symmetry axis is $\tau_{\rm V} = 0.3$ for
the V-band, its half-opening angle equals $\theta_C = 30\degr$.
Whilst in \citet{goosmann2007} we assume a dust composition
reproducing the extinction in our Galaxy, we here apply a different
dust parameterization, that inferred from AGN extinction curves
\citep{gaskell2004,gaskell2007}, which show a flatter far-UV
slope than those for our Galaxy. Our AGN dust model implemented here
contains 85\% ``astronomical silicate'' and 15\% graphite. The
distribution, $n(a)$, of grain radii, $a$, obeys to the power-law
$n(a) \propto a^{-2.05}$. The resulting polarization and flux spectra,
normalized to the central illuminating flux, are shown in
Fig.~\ref{fig:res} (left).

Comparison to the results in \citet{goosmann2007} reveals some clear
differences: The polarization spectra are weaker at shorter wavelengths than
for the Galactic dust model. Around 4000 \AA, the polarization curves drop
down to nearly zero and then they increase again steeply to longer
wavelengths. For Galactic dust the polarization spectra show a more gradual
increase over the whole wavelength range considered. Because of the larger
average grain size, the scattering efficiency of the AGN dust is significantly
smaller. The flux seen beyond the edge of the cone, for $i > \theta_C$, is
therefore low.

We find similar trends for a larger half-opening angle of $\theta_C =
45\degr$, as shown in Fig.~\ref{fig:res} (right). However, the
obtained polarization percentages are lower, than for the narrower
cone, because the observer detects the integration of a broader range
of polarization vectors. The scattered flux at $i > 45\degr$, on the
other hand, is higher due to higher coverage of the central source.

\end{document}